\documentclass[runningheads]{llncs}
\usepackage[T1]{fontenc}
\usepackage{graphicx}
\usepackage[hidelinks]{hyperref}
\begin{document}
\title{Variable Stiffness \& Dynamic Force Sensor for Tissue Palpation}
\author{Abu Bakar Dawood\inst{1}\orcidID{0000-0002-1800-7919} \and Zhenyu Zhang\inst{2}\orcidID{0000-0003-0797-5540} \and Martin Angelmahr\inst{2}\orcidID{0000-0002-3147-8918}
Alberto~Arezzo\inst{3}\orcidID{0000-0002-2110-4082} \and
Kaspar Althoefer\inst{1}\orcidID{0000-0002-1141-9996
}}
\authorrunning{Dawood et al.}
\institute{Centre for Advanced Robotics @ Queen Mary, Queen Mary University of London, United Kingdom \and Fraunhofer Institute for Telecommunications, Heinrich Hertz Institute, Goslar, Germany \and
Department of Surgical Sciences, University of Torino, Italy\\
\email{a.dawood@qmul.ac.uk}}
\maketitle              % typeset the header of the contribution
\begin{abstract}
Palpation of human tissue during Minimally Invasive Surgery is hampered due to restricted access. In this extended abstract, we present a variable stiffness and dynamic force range sensor that has the potential to address  this challenge. The sensor utilises light reflection to estimate sensor deformation, and from this, the force applied. Experimental testing at different pressures (0, 0.5 and 1 PSI) shows that  stiffness and force range increases with pressure. The force calibration results when compared with measured forces produced an average RMSE of 0.016, 0.0715 and 0.1284 N respectively, for these pressures.

\end{abstract}
\section{Introduction}
% \subsection{A Subsection Sample}
Palpation is a key technique in which clinicians use their hands and fingers to localise and identify tissue abnormalities. By exerting force upon human tissue, clinicians can glean information about tissue condition during open surgery \cite{konstantinova2014implementation}. However, in Minimally Invasive Surgery (MIS)  conventional palpation cannot be used, making tissue examination difficult \cite{konstantinova2017palpation}. Force sensing probes, adapted for minimally invasive surgery, can be used to detect tissue abnormalities such as tumors, which are significantly stiffer than healthy tissue by a factor of ten) \cite{ahn2010mechanical}. 
\smallskip 

Numerous solutions have been proposed to compensate for the lack of haptic feedback in Minimally Invasive Surgery. These solutions leverage sensing technologies including optical \cite{Ahmadi2010} \cite{Dawood_optical_2022} \cite{Xie2013}, capacitive \cite{Dawood_Taros_cap_2020} \cite{Dawood_cap_2023} \cite{Dawood_cap_IROS_2020}, vibro-acoustic \cite{Suhn2023} and pneumatic \cite{indika2013}. Typically, force sensors have a predetermined force sensing range and sensitivity - indeed only a few attempts have been made to develop sensors with adjustable force range and sensitivity. Raitt et. al. \cite{Raitt2022} developed a stiffness controllable sensing tip that uses a camera to observe the deformation of a silicone membrane. The membrane stiffness can be controlled by pneumatic pressure, resulting in an adjustable force range. Another dynamic force range sensor ESPRESS.0 developed by Jenkinson et. al. also uses pneumatic pressure to adjust the stiffness of the membrane, although in this
instance, the force is measured with a camera, by tracking
the fluid, coupled with the membrane, inside a tube \cite{Jenkinson2023}.

This extended abstract introduces a soft force sensor whose stiffness can be adjusted, that employs light reflection to estimate deformation and, from that, the force. Its adjustable stiffness, which is controlled by pneumatic input, means that it can function over a variable force range. Experiments were conducted at different internal pressures and the calculated force was compared to the measured force.

\section{Materials and Methods}

To fabricate the sensor, moulds were designed using Solidworks and 3D printed using Ultimaker S3. Polylactic Acid (PLA) was used to 3D print the moulds. EcoFlex 00-50 parts A \& B were mixed in equal quantities along with black dye. This black silicone mixture was then degassed using a vacuum chamber. A plastic fibre was placed inside the mould for  integration into the silicone before closing the mould. This fibre would restrict the ballooning of the silicone dome, preventing it from rupture. EcoFlex was poured into the mould and degassed again to remove any air pockets within the mould. The EcoFlex was then cured at room temperature. 

\smallskip 

\begin{figure}[h!]
\centering
\includegraphics[width=0.9\columnwidth]{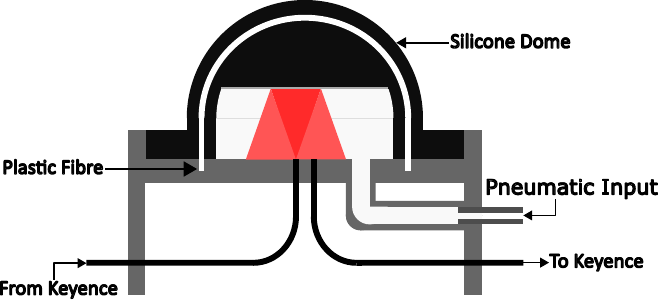}
\caption{Our proposed palpation sensor - a black silicone dome with an integrated plastic fibre to restrict ballooning.}
\vspace{-6mm}
\label{fig:cross_section}
\end{figure}

The Silicone dome, 40mm in diameter, was then taken out of the mould and Aluminum powder was brushed onto its inner side to enhance its reflectivity. A holder for the dome was 3D printed with a pneumatic channel and holes for both emitting and receiving optical fibres, as well as for the integrated plastic fibre. The dome was glued to the holder, ensuring no air leakage. The cross-sectional view of the sensor is shown in Figure{~\ref{fig:cross_section}}.

% \smallskip 

The sensor was pressurised using a syringe pump, which was  connected to a stepper motor and controlled by an Arduino controller. The air pressure inside the dome was measured using a pressure sensing IC (NPA-500B-005D by Amphenol Advanced Sensors).

\begin{figure}[h]
\centering
\includegraphics[width=0.95\columnwidth]{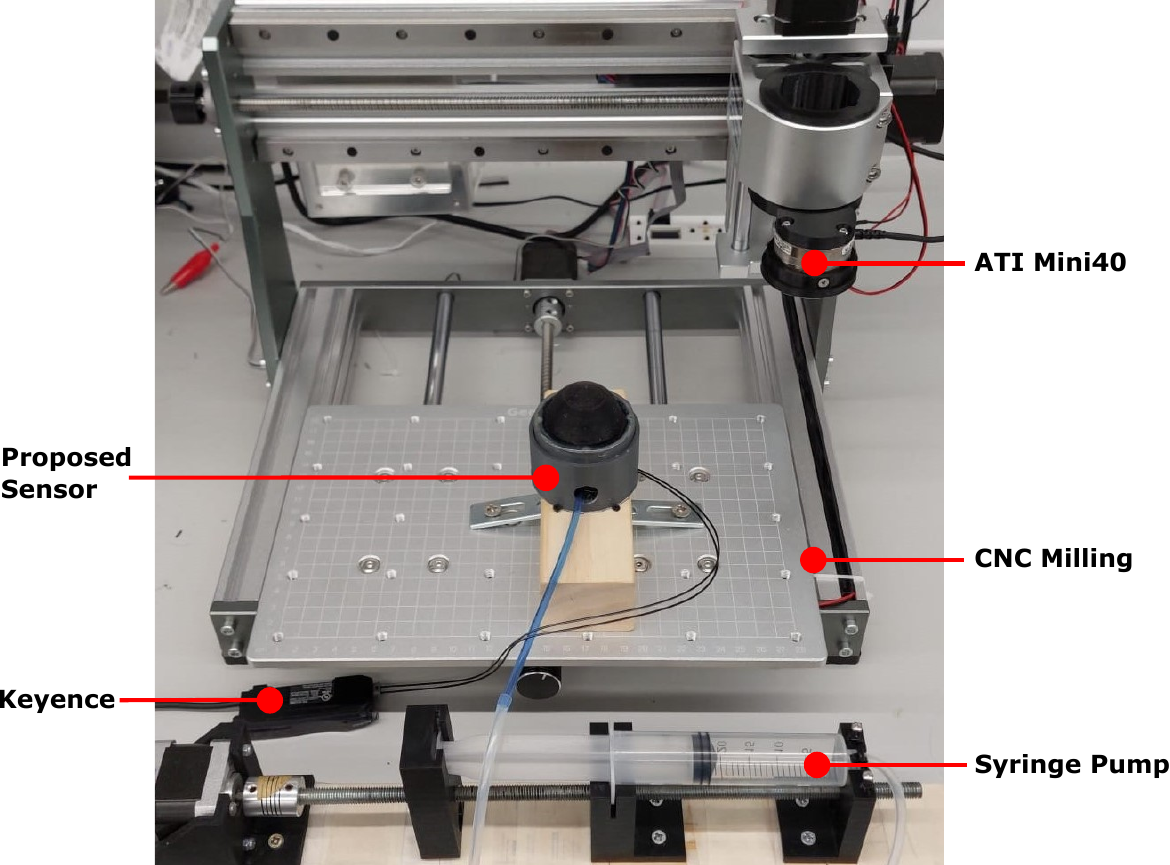}
\caption{The experimental setup showing a modified CNC Milling machine with an ATI Mini40, Syringe pump and Keyence.}
\label{fig:experimental_setup}
\end{figure}

\begin{figure*}[h!]

\includegraphics[width=0.55\linewidth]{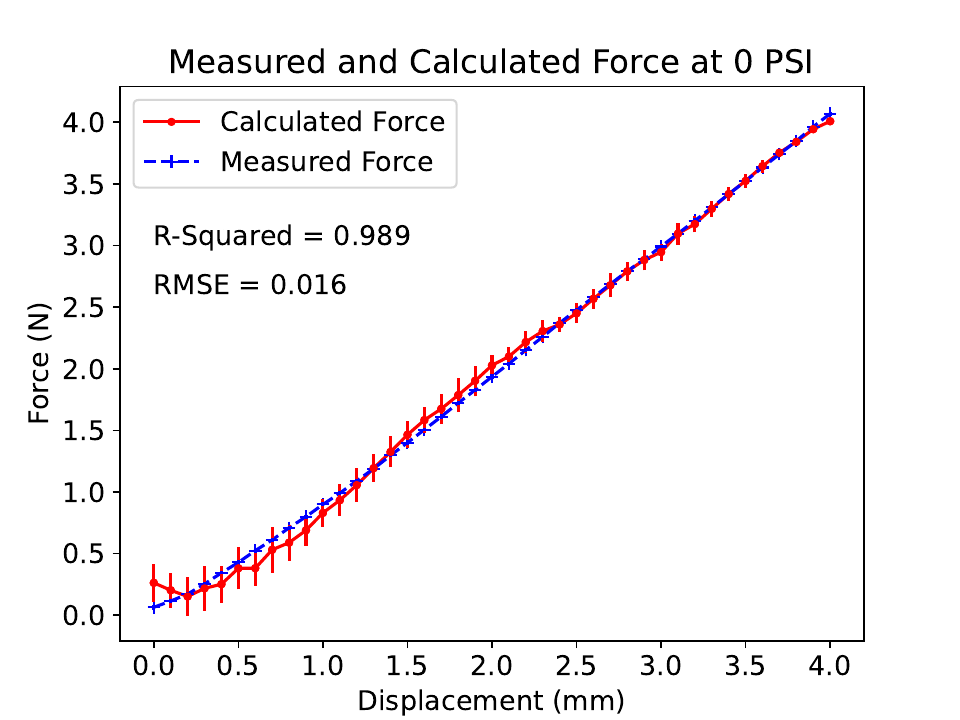}
\hspace*{-0.6cm}\includegraphics[width=0.55\linewidth]{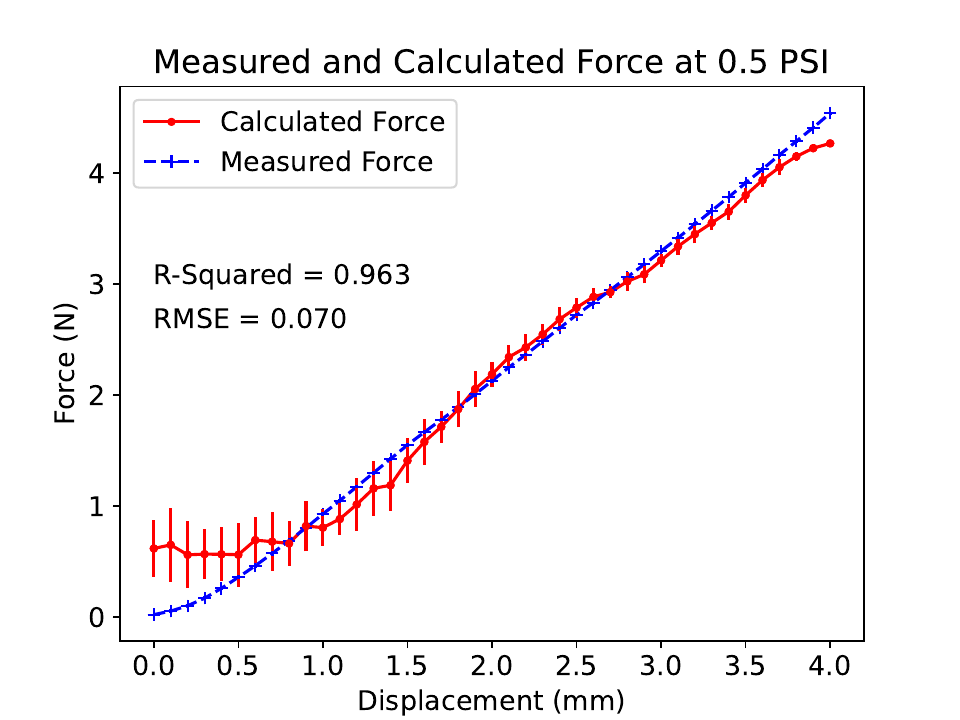}
\hspace*{3cm}\includegraphics[width=0.55 \linewidth]{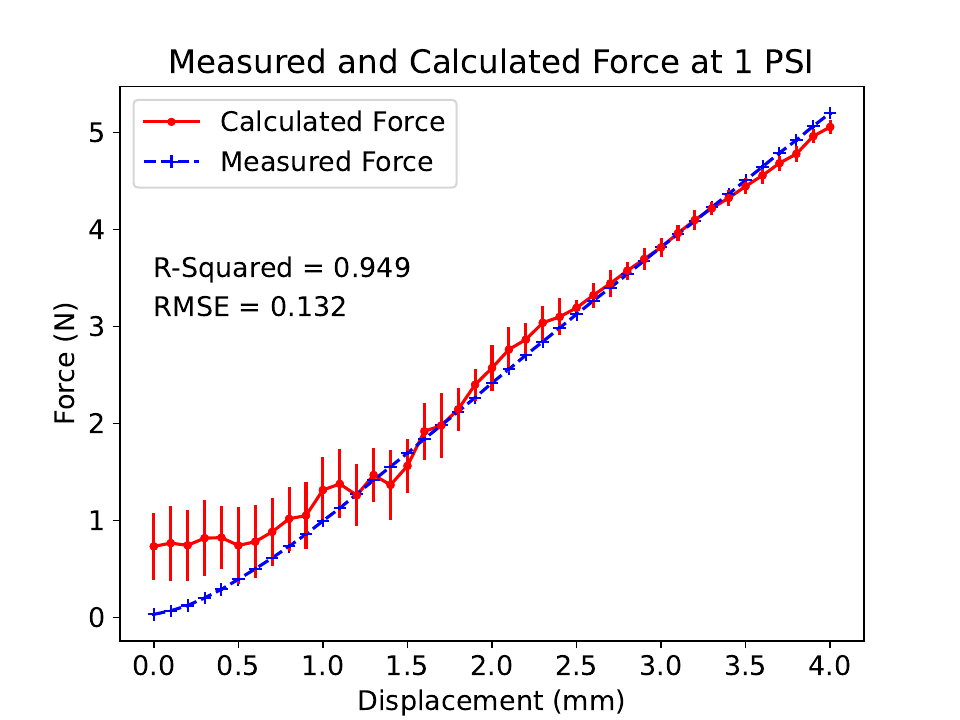}
\caption{Calculated and measured forces by force torque sensor at 0, 0.5 and 1 PSI, respectively. Error bars show the standard deviation. R-squared and RMSE between the calculated values and measured values are also shown.}

\vspace{-4mm}
\label{fig:plots}

\end{figure*}

The emitting and receiving optical fibres were connected to an optoelectronic system (Keyence). A Keyence emits light and measures the intensity of the light it then receives, converting it to a voltage signal which in our set-up was measured by the Arduino. A modified CNC milling machine (Genmitsu 3018-PRO) with an ATI Mini40 force and torque sensor was used to test the sensor. A python program was developed to control the movement of the CNC machine, communicate with the Arduino, and for data acquisition and data recording. The experimental setup is shown in Figure~\ref{fig:experimental_setup}.

The sensor was tested at 3 different pressures 0 PSI (0 Pa), 0.5 PSI (3.447 kPa) and 1 PSI (6.894 kPa). The sensor was indented using a flat indenter which was displaced by 4mm, and each experiment was repeated 3 times. For each pressure, the stiffness of the sensor was calculated by using Hooke's law,\vspace{-1mm}
\begin{equation}
F = kx
\vspace{-1mm}
\end{equation}

where $F$ is the applied force, $k$ is the stiffness of the spring and $x$ is the displacement. To find the stiffness of our sensor, the force measured by the ATI Mini40 sensor under a displacement of 4mm was used.

\vspace{2mm}

For one sample of each internal pressure, the data from the optical proximity sensor was then calibrated for distance, by using a 4\textsuperscript{th} order polynomial. This calculated distance was then multiplied by the stiffness, to estimate the force value.

\section{Results}

Sensor stiffness values, calculated by using ground truth values of displacement and force, for 0, 0.5 and 1 PSI were $1016.13\pm 1.29 N/m$, $1134.34\pm 7.13 N/m$ and $1301.633\pm 1.21 N/m$, respectively. The equations obtained after the polynomial fitting, for the calculation of displacement ($y$) from the optical data ($x$), are:\\

\vspace{0mm}
At 0 PSI:
\begin{equation}
    y= -43.328x^4+ 4.367\times10^{2}x^3
-1.6605\times10^3x^2\\+2.8294\times10^3x -1.82325\times10^3
\end{equation}

% \end{multline}

At 0.5 PSI:
\begin{equation}
y= -6.352\times10^{2}x^4+ 6.0185\times10^{3}x^3\\
-2.1382\times10^4x^2+3.3765\times10^4x -1.99978\times10^4
\end{equation}

At 1 PSI:
\begin{equation}
y= -3.2323\times10^{2}x^4+ 3.0106\times10^{3}x^3\\
-1.0519\times10^4x^2+1.6346\times10^4x -9.53143\times10^3
\end{equation}

The calculated value of displacement in meters when multiplied by the stiffness, resulted in the calculated force. The two metrics used for the comparison of calculated and measured force are R-squared and Root Mean Squared Error, shown in the Table~\ref{tab:classif_results}. We note that that curve fitting was performed only on Dataset 3 for each of the internal pressures.

\begin{table}[h]
\centering

\caption{R-Squared and Root Mean Squared Error (RMSE) of all the three datasets at three pressures.}
\resizebox{\columnwidth}{!}{%
\begin{tabular}{|c||c|c||c|c||c|c|}
\hline
 & \multicolumn{2}{c||}{Data 1} & \multicolumn{2}{c||}{Data 2} & \multicolumn{2}{c|}{Data 3} \\ \cline{2-7} 
    & RMSE & $R^2$ & RMSE & $R^2$ & RMSE & $R^2$ \\ \hline
    0 PSI & 0.016 & 0.989 & 0.016 & 0.989 & 0.016 & 0.989    \\ \hline
    0.5 PSI & 0.0697 & 0.9637 & 0.0746 & 0.9604 & 0.0702 & 0.9633   \\ \hline
    1 PSI & 0.1316 & 0.949 & 0.1286 & 0.9506 & 0.125 & 0.9523   \\ \hline
    
\end{tabular}
}
\label{tab:classif_results}
\end{table}

\section*{DISCUSSION}

The stiffness of the sensor, calculated by using the ATI Mini40 and the displacement of the machine, shows a direct correlation with the applied pressure. It appears that by increasing the applied internal pressure, the force range of the sensor increases while the sensitivity decreases. This can also be observed in Figure~\ref{fig:plots} - as the pressure increases, the standard deviation of the calculated force, especially for lower forces, also increases.

\vspace{2mm}

Table~\ref{tab:classif_results} shows the R-squared and Root Mean Squared Error of measured force and calculated force, for all three datasets. Both the metrics are very consistent for force estimation at 0 PSI and Figure~\ref{fig:plots} confirms this. The metrics start showing deviation from previous behaviour with lower R-squared values and higher RMSE, as the pressure increases. The least favorable metrics are associated with dataset 1 at 1PSI, with an RMSE of 0.1316 N and an R-squared value of 0.949.

\vspace{2mm}

We plan to employ data driven methods to estimate interaction forces and the stiffness of palpated soft tissue. Future work will also focus on miniaturising the sensor for use in MIS, which will mean aiming for maximum diameters of 10mm so as to enable them to fit through trocar ports.

%
% the environments 'definition', 'lemma', 'proposition', 'corollary',
% 'remark', and 'example' are defined in the LLNCS documentclass as well.
%

\begin{credits}
\subsubsection{\ackname}

This work was funded by UK Research and Innovation (UKRI) under the UK government's Horizon Europe funding guarantee [grant \# N°101092518] and funded by the European Union. \\
The design files, dataset, program and the instructions are uploaded to a Github repository 
\href{https://github.com/abubakardawood/variable_stiffness_sensor_1.0}{\textit{\textbf{Variable Stiffness Sensor}}}.

\end{credits}
\bibliographystyle{unsrt}
\bibliography{references}




% LNCS Homepage, \url{http://www.springer.com/lncs}, last accessed 2023/10/25
% \end{thebibliography}

\end{document}